# High Electron Mobility in Vacuum and Ambient for PDIF-CN$_2$ Single-Crystal Transistors

Anna S. Molinari*[1], Helena Alves[2], Zhihua Chen[3], Antonio Facchetti*[3] and Alberto F. Morpurgo*[4]

[1]Kavli Institue of Nanoscience, Delft University, Lorentzweg 1, 2628CJ, Delft, (the Netherlands); [2]INESC-MN, Rua Alves Redol 9, 1000-029 Lisboa ( Portugal);[3]Polyera Corporation, 8025 Lamon Avenue, Skokie, IL 60077 (USA); [4]DPMC and GAP, University of Geneva, 24 Quai Ernest-Ansermert CH-1211, Geneva (Switzerland)

Semiconductor devices based on organic molecules are of interests for large area electronics as well as to underscore fundamental charge transport effects in molecular solids.[1] The field-effect transistor (FET, Fig. 1) is an electronic device where the current between the source and drain contacts ($I_D$), for a given source-drain voltage ($V_{DS}$), is modulated by the application of the gate-source bias ($V_{GS}$). In this device one of the key figures of merit is the field-effect mobility ($\mu_{FET}$), which should be as large as possible to enable the fabrication of devices with greater performance.[1] Using the FET architecture, the carrier mobilities of several polycrystalline/amorphous p-channel (charge = hole) and n-channel (charge = electron) organic semiconductor films have been tested.[2] Although FETs based on polycrystalline organic semiconductor will probably be the first to be implemented into a commercial product,[1] FETs based on single-crystal organic semiconductors are far more interesting to probe ultimate charge transport efficiencies.[3] To date, among the greatest hole mobility values have been reported for single-crystal FETs (SCFETs) based on tetracene ($\mu_{FET}$ ~ 2.4 cm$^2$/Vs, PDMS gate dielectric)[3b] and rubrene ($\mu_{FET}$ ~ 20 cm$^2$/Vs, air-gap gate dielectric; $\mu_{FET}$ < 10 cm$^2$/Vs, other gate dielectrics).[3c,d,l] Other p-channel single crystal semiconductors such as anthracene, pentacene, TIPS-pentacene, DPh-BDSe, and BPT2 exhibit $\mu_{FET}$s = 0.02-2.2 cm$^2$/Vs.[3e-j] As in the case of thin-film-based FETs, single-crystal FETs based on n-channel semiconductor are far more rare and fall short in performance when compared to p-channel semiconductors. The best n-channel SCFETs reported to date were fabricated with copper perfluorophthalocyanine (FCuPc, $\mu_{FET}$ = 0.2 cm$^2$/Vs, parylene gate dielectric)[3k] and tetracyanoquinodimethane (TCNQ, $\mu_{FET}$ = 1.6 cm$^2$/Vs in vacuum, air-gap gate dielectric).[3c] Despite the overall lower performance of thin-film as compared to single-crystal FETs, several n-channel semiconductors have shown far larger performance than FCuPc and TCNQ in thin-film-based FETs.[2d-g,4] This suggests that through an appropriate choice of the molecular material, n-channel SCFETs may approach –or possibly even exceed- the $\mu_{FET}$ of the best p-channel transistors.

Among the n-channel semiconductors for FETs, *N,N*'-bis(n-alkyl)-(1,7&1,6)-dicyanoperylene-3,4:9,10-bis(dicarboximide)s (PDIR-CN$_2$), have shown great potential.[5] Some PDIR-CN$_2$ derivatives exhibit excellent electrical performance and remarkable environmental stability ($\mu_{FET}$ ~ 0.01-0.1 cm$^2$/Vs for PDI8-CN$_2$ where R = C$_8$H$_{17}$; $\mu_{FET}$ ~ 0.1-0.6 cm$^2$/Vs for PDIF-CN$_2$ where R = CH$_2$C$_3$F$_7$) for vapor-/solution-deposited thin-film FETs.[5] Since it is now established that the greater structural order of single-crystalline materials enables high-quality devices with many different molecular materials,[6] we decided to investigate single-crystal transistors based on PDIF-CN$_2$ which exhibits the largest thin-film $\mu_{FET}$ within the core-cyanated perylene family. Thanks to the *N*-fluorocarbon functionalization, PDIF-CN$_2$ is thermally stable and sublimes quantitatively, thus enabling high-quality single crystal growth by vapor phase transport. Furthermore, PDIF-CN$_2$ crystal structure is known.[5b] Here we demonstrate that PDIF-CN$_2$-based SCFETs with PMMA as the gate dielectric exhibit ~ 10x larger mobility than the corresponding thin-film devices both in the air and in vacuum. Furthermore, other PDIF-CN$_2$ SCFET characteristics are also promising and include near-zero threshold voltage ($V_{th}$) and sub-threshold slope (*S*) and current on/off ratio ($I_{on}:I_{off}$) of the same order of the very best p-channel single-crystal devices.

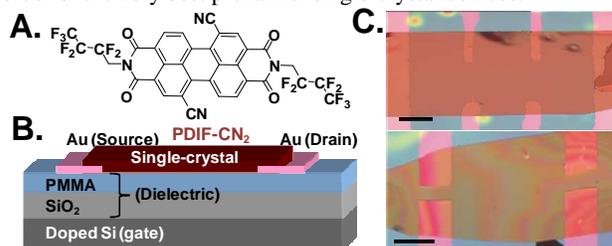

**Figure 1.** A. Molecular structure of PDIF-CN$_2$. B. Schematic layout of a single-crystal field-effect transistor. C. Optical microscope images of PDIF-CN$_2$ devices used in our investigations. The bar is 200 μm.

Single-crystal FETs (Fig. 1) were fabricated according to the procedure previously reported for p-channel semiconductors.[6a] PDIF-CN$_2$ (Polyera ActivInk N1100) red rectangular crystals with lengths of few millimetres and widths up to 500 μm were obtained by physical vapour transport in a stream of Ar.[7] These thin crystals (<1 μm thick) are difficult to handle but we were able to laminate them onto heavily doped Si(gate)/SiO$_2$-PMMA (dielectric) substrates with prefabricated Ti/Au (source-drain) contacts. This bottom-up approach avoids damaging the delicate surface of the organic material. The bilayer dielectric structure was used to minimize electron trapping by the hydroxyl groups of the SiO$_2$ surface.[8] Both two- and four-terminal configurations were employed to account for the possible influence of the contact resistance. Figures 1 and S1 provides optical microscope images of some single crystal devices.

Typical electrical characteristics of PDIF-CN$_2$ SCFETs measured in vacuum and in air are shown in Figure 2. One of the first striking features is that they are virtually free of hysteresis and in most cases the threshold voltage is between -5V and +5V usually close to 0 V. This value is much smaller than that measured for vapor-deposited PDIF-CN$_2$ thin-film transistors ($V_{th}$ = -20 ~ -30 V)[5b] demonstrating that the dopant concentration in PDIF-CN$_2$ single-crystals is substantially reduced as compared to thin-films. This lower dopant concentration may be due to the enhanced chemical purity of the vapor-grown single-crystals. Another device feature that should be stressed is the linear behavior of the $I_D$ at low $V_{DS}$ (Fig 2A), demonstrating the absence of strong contact effects. This result is remarkable considering that high work-function gold is used as the metal contact. Low workfunction metals are usually required to inject electrons in n-channel organic devices.[9]

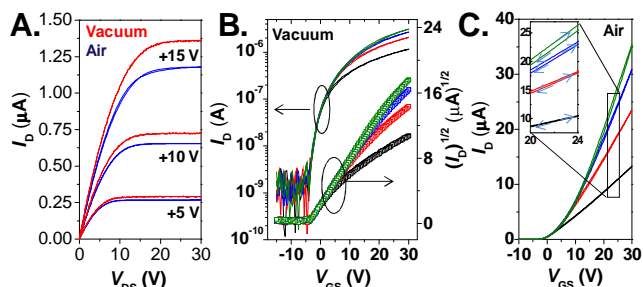

**Figure 2.** A. Output plots for the same PDIF-CN$_2$ FET measured in vacuum (red) and in ambient (blue). B. Transfer plot of the $I_D$ versus $V_{GS}$ measured in vacuum for different values of the $V_{DS}$ (+5V=black, +10V=red, +15V=blue, and +20V=green). The corresponding square root of $I_D$ values are also shown in the same plot. C. Transfer plot of the $I_D$ versus $V_{GS}$ measured in air for different values of the $V_{DS}$ (+5V=black, +10V=red, +15V=blue, and +20V=green). The inset shows the negligible I-V hysteresis. For this device the crystal width W is 680 µm and the channel length L ~1mm.

PDIF-CN$_2$ electron mobility for several single-crystal devices was extracted in the linear regime according to the equation $\mu_{FET} = L/WC_iV(\partial I_D/\partial V_{GS})$, where $V$ is the potential difference measured between the voltage probes, $L$ their distance, $W$ is the channel width, and $C_i$ is the measured capacitance-per-unit-area between gate and conduction channel. Figure 3 shows the $\mu_{FET}$ and the $I_{on}$:$I_{off}$ for several devices measured both in vacuum and in ambient conditions. The vacuum $\mu_{FET}$ values range from ~6 to 1 cm$^2$/Vs whereas the mobility in ambient ranges between ~3 and 0.8 cm$^2$/Vs. Note that similar values were obtained for both two- and four- terminal measurement configurations confirming the good quality of the contacts (interestingly, a comparable mobility reduction from vacuum to ambient, ~ 2×, has recently been observed for TFTs based on solution-processed/vapor-deposited PDIF-CN$_2$ polycrystalline)[10]. However, the values obtained for single-crystals are the largest mobility values reported to date (in ambient and in vacuum). The best vacuum mobility is ~ 4× larger than the best reported for TCNQ in vacuum.[3] Such mobility value is close/larger to that of holes in rubrene SCFETs ($\mu_{FET}$ = 0.01 - 10 cm$^2$/Vs)[8,31] in devices fabricated using a gate dielectric different than air-gap ($k > 1$). Other FET parameters such as $S$ and the $I_{on}$:$I_{off}$ were analyzed. In our PDIF-CN$_2$ devices $S$ = ~1.9 to 5 V decade$^{-1}$, which is comparable to those of tetracene but larger than the best single-crystal rubrene FETs when the gate dielectric capacitance is taken into account (see Supporting Information).[11] Finally, the $I_{on}$:$I_{off}$ of PDIF-CN$_2$ SCFETs is ~ 10$^3$/10$^4$ for different devices, comparable/slightly lower than in high-quality p-channel SCFETs, but larger than that of thin-film PDIF-CN$_2$ FETs.[5b]

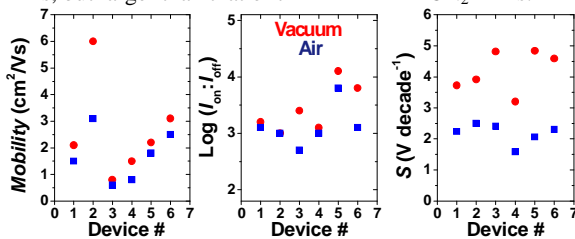

**Figure 3.** Electron mobilities (four-probe), $I_{on}$:$I_{off}$, and $S$ for six single-crystal PDIF-CN$_2$ FETs measured in vacuum (red) and in the air (blue).

The difference between the PDIF-CN$_2$ thin-film (~0.1 − 0.6 cm$^2$/Vs) and single-crystal (~1 − 6 cm$^2$/Vs) mobility values may have several origins. Efficient charge transport in organic semiconductors requires an uniform and continuous film morphology; however vapour-/solution-deposited thin films are usually characterised by grain boundaries and random crystallite molecular orientation which limit the device performance. A single crystal represents an almost ideal situation for charge transport characterized by absence of macroscopic grain boundaries, a very smooth and homogeneous surface, and micrometer-extended ordered molecular arrangement. Moreover, the crystal growth process provides an additional purification step reducing charge traps.

In conclusion, we demonstrated n-channel SCFETs exhibiting field-effect mobility values approaching those of the best p-channel FETs when fabricated with dielectric materials with comparable $k$ values. We believe that these data will provide additional stimuli for the synthesis and development of chemically-engineered n-channel organic semiconductors. Furthermore, it will be interesting to study how this single-crystal semiconductor performs with unique dielectric materials and device architectures.[12]

**Acknowledgements.** We are grateful to NanoNed and NWO for financial support. HA also acknowledges FCT for financial support under contract nr. SFRH/BPD/34333/2006.

**Supporting Information Available**: Single-crystal device fabrication and measurements details. This material is available free of charge via the internet at http://pubs.acs.org.

# Supporting Information

## High Electron Mobility in Vacuum and Ambient for PDIF-CN$_2$ Single-Crystal Transistors

Anna S. Molinari*, Helena Alves, Zhihua Chen, Antonio Facchetti* and Alberto F. Morpurgo*

*Kavli Institue of Nanoscience, Delft University, Lorentzweg 1, 2628CJ, Delft, the Netherlands;INESC -MN, Rua Alves Redol 9, 1000-029 Lisboa ( Portugal); Department of Condensed Matter Physics, Ecole de Physique 24 Quai Ernest-Ansermet CH-1211, Geneva (Switzerland) and Polyera Corporation, 8025 Lamon Avenue, Skokie, IL 60077 (USA)*

Single-crystal field-effect transistors (FETs) based on a fluorocarbon-substituted dicyanoperylene-3,4:9,10-bis(dicarboximide) [**PDIF-CN$_2$**] were fabricated by lamination of the semiconductor crystal on Si-SiO$_2$/PMMA-Au gate-dielectric-contact substrates. These devices were characterized both in vacuum and in the air and exhibit electron mobilities of ~ 6-3 cm$^2$V$^{-1}$s$^{-1}$ and ~ 3-1 cm$^2$V$^{-1}$s$^{-1}$, respectively, $I_{on}$:$I_{off}$ > 10$^3$, and near-zero threshold voltage.

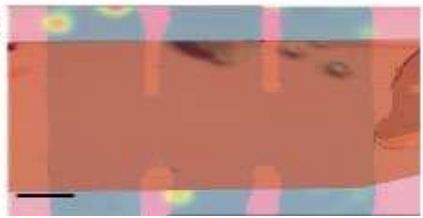

**Device Fabrication**

PDIF-CN$_2$ is commercially available from Polyera Corporation (Polyera Activink™ N1100). Red rectangular crystals were obtained by physical vapour transport in a stream of argon gas, typically with a length of few millimetres and width up to hundreds of μm (> 500 um). Single crystal FET devices were assembled in a similar technique used before for p-type semiconductors. It consists in laminating a thin crystal (<1 μm) onto a heavily doped silicon substrate (the gate) covered by SiO$_2$ (200 nm thick) (the dielectric) with prefabricated source-drain contacts. This bottom-up approach avoids damaging the delicate surface of the organic material. In PDIF-CN$_2$ single crystal OFETS a layer of polymethyl-methacrylate (PMMA; ~ 100 nm thick) was interpose between the semiconductor and the SiO$_2$. The extra PMMA dielectric layer was used to avoid electron traps generated by hydroxyl groups on the SiO$_2$ surface. Figure 1 provides a schematic illustration of the OFET, with a structure of Si/SiO$_2$/PMMA/Ti/Au/PDIF-CN$_2$, as well as optical microscope images of some single crystal devices. Source and drain electrodes consisting of 5 nm Ti and 20 nm Au were deposited in high vacuum (10$^{-7}$ mbar) by electron-beam evaporation through a hand-made shadow mask. A four terminal configuration was employed, to prevent that a large, unwanted contact resistance, affected the measurement of the transistor channel. All electrical measurements were performed both in vacuum and in air using an Agilent E5270A or a HP 4192A parameter analyzer.

The FETs have a channel length (*L*) of 1mm and width (*W*) depending on the width of the crystal (usually hundreds of micrometers). The mobility was calculated in the linear transport regime in a four point configuration, using the equation:

$$\mu = \frac{L}{WC_iV}\frac{\partial I_D}{\partial V_{GS}} \qquad (1)$$

where $V$ is the potential difference measured between the voltage probes, $L$ their distance, $W$ is the channel width, and $C_i$ is the measured capacitance-per-unit-area between gate and conduction channel. $V_{GS}$ is the gate voltage and $I_D$ the drain-source current. The total capacitance-per-unit area of the SiO$_2$-PMMA stack was measured. The value found was ~ 8 nF/cm$^2$, which is consistent with the known dielectric constants of the two materials.

The sub-threshold slope was also considered. It quantifies the change in input voltage ($V_{GS}$) that is required to change the output current ($I_D$) by one order of magnitude when the transistor is biased in the subthreshold regime and is defined by:

$$S \equiv \frac{dV_{GS}}{d(\log I_D)} \quad (2)$$

$S$ is mainly determined by the quality of the insulator/semiconductor interface. Specifically, if the amount of disorder is small then $I_D$ has a fast increase with increasing $V_{GS}$, resulting in a lower $S$ value (i.e., low $S$-values correspond to better performance). Quantitatively, the value of $S$ can be readily determined from the plot of $log(I_D)$-vs-$V_{GS}$, but it was also normalized to the capacitance of the insulating layer $C_i$:

$$\tilde{S} \equiv \frac{dV_{GS}}{d(\log I_D)} C_i \quad (3)$$

This, usually, allows comparison between devices, avoiding different FET characteristics like the dielectric thickness. From the provided parameters Ci = 8 nF/cm$^2$, and $\tilde{S}$ = 15 to 40 V nF dec$^{-1}$ cm$^{-2}$, which is comparable to tetracene (28 VnF decade$^{-1}$cm$^{-2}$) but larger than the best single-crystal rubrene FETs.

Finally, the ON/OFF current ratio was measured from the plot of $log(I_{SD})$-vs-$V_G$.

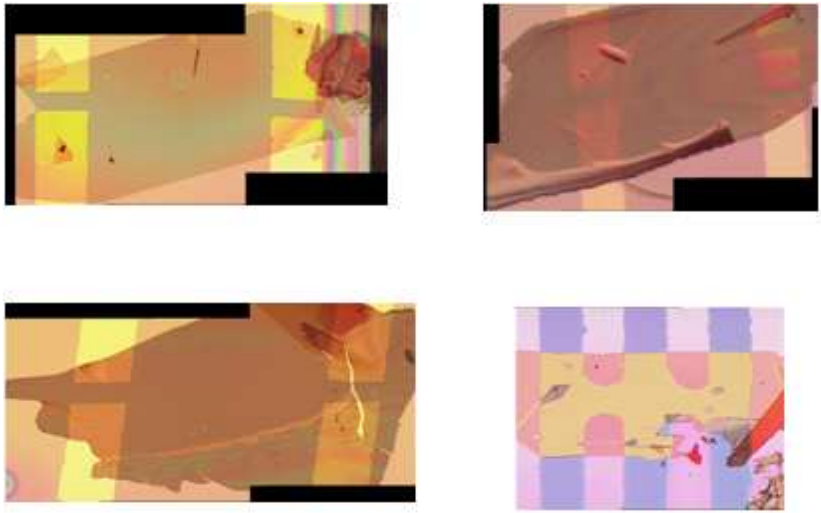

*Figure S1.* Optical images of some of the single-crystal TFTs fabricated with **PDIF-CN$_2$**